\documentstyle[preprint,prc,eqsecnum,aps,epsf]{revtex}

\tightenlines   

\begin{document}
\draft

\def\lsim{\thinspace{\hbox to 8pt{\raise -5pt\hbox{$\sim$}\hss{$<$}}}\thinspace}
\def\rsim{\thinspace{\hbox to 8pt{\raise -5pt\hbox{$\sim$}\hss{$>$}}}\thinspace}

\title{Off-Shell Structures of Nucleon-Nucleon t-matrices and their
Influence on Nucleon-Nucleus Elastic Scattering Observables}

\author{S.P.~Weppner}
\address{Natural Arts and Sciences, Eckerd College, St. Petersburg, FL 33711} 

\author{Ch.~Elster}
\address{
Institute of Nuclear and Particle Physics,  and
Department of Physics and Astronomy, \\ Ohio University, Athens, OH 45701}

\author{ D. H\"uber}
\address{
Theoretical Division, Los Alamos National Laboratory, 
Los Alamos NM, 87545 }

\date{\today}

\maketitle
 
\begin{abstract}
The sensitivity of nucleon-nucleus elastic scattering observables
to the off-shell structure of nucleon-nucleon t-matrices, derived
from realistic NN potentials, is investigated within the context
of a full-folding model based on the impulse approximation.
Our study uses recently developed NN potential models, which describe
a subset of the NN data base with a $\chi^2$ per datum $\sim$1, which
means that the NN t-matrices are essentially on-shell equivalent.
We calculate proton-nucleus elastic scattering observables for
$^{16}$O, $^{40}$Ca, and $^{208}$Pb between 100 and 200~MeV laboratory
energy. We find that the elastic scattering observables are insensitive
to off-shell differences of the employed NN t-matrices. A more detailed
investigation of the scattering equation and the optical potential
as given in a factorized approximation reveals that the elastic 
scattering observables do not sample the NN t-matrices very far
off-shell where they exhibit differences. 

\end{abstract}

\vspace{5mm}

\pacs{PACS: 25.40.Cm, 24.10-Ht}

\pagebreak


\narrowtext

\section{Introduction}

\hspace*{10mm}
Theoretical investigations of  nucleon-nucleon (NN) transition
amplitudes 
in their off-shell domain have a long history in the study
of few  and many-nucleon systems. Often those investigations 
were inconclusive due to the lack of NN potentials which
describe the NN observables with equally high accuracy. Current interest
in this issue is driven by the recent development of NN potentials which
below pion production threshold describe the NN data base with a
$\chi^2$ per datum $\sim$ 1 \cite{nijm,cdbonn,argonne}. Transition
amplitudes derived from these
potentials can be considered on-shell equivalent. Their
different theoretical derivation gives rise to different off-shell
extrapolations.

\hspace*{10mm}
At intermediate energies elastic nucleon-nucleus (NA) scattering can be
successfully described by the leading term in the spectator expansion
of multiple scattering theory \cite{Corr,Sicil,med2}. Here an optical 
potential is derived, which in its most general form is given by the
expectation value of the NN transition amplitude and the ground state
of the target nucleus. This `full-folding' optical potential involves
the convolution of the fully off-shell NN scattering amplitude with a 
realistic single particle nuclear density matrix.

\hspace*{10mm}
Recently, significant advances have been made in accurately handling
these off-shell degrees of freedom in elastic NA scattering 
\cite{hugo1,hugo2,FFC,ff1,edff}.
Those studies have demonstrated that an accurate treatment of the
off-shell structure of the NN transition amplitude is needed for a
proper account of the theory. In order to cleanly isolate if NA elastic
scattering observables are sensitive to different off-shell structures
of realistic NN transition amplitudes, it is necessary to start from NN
potentials which describe the NN data base with a high degree of
accuracy. Our present study is based on the potential models for the
NN interaction recently developed by
the Nijmegen group \cite{nijm} and the charge-dependent Bonn (CD-Bonn) 
potential \cite{cdbonn}. With NN transition amplitudes derived from these
potentials we calculate full-folding optical potentials and elastic
scattering observables  for proton scattering from a variety of nuclei in
the energy regime between 100 and 200 MeV projectile energy. Although
the off-shell structure of the NN t-matrices is an important ingredient
in the calculations, we find that off-shell differences between 
the models are not discernible by NA elastic scattering.

\hspace*{10mm}
In order to understand this result and obtain more insight
 which regions of the off-shell NN t-matrix are
sampled in a calculation of NA elastic scattering observables, we 
use the optimum factorized or off-shell $t\rho$ formulation of the
optical potential. This formulation, quite a good approximation
in the energy regime around 200 MeV and higher, has the
advantage that the fully off-shell NN t-matrix enters  together with
an on-shell density.

\hspace*{10mm}
The structure of this article is as follows. First we review in Section~II 
the relevant expressions for the full-folding optical potential as used
in our calculations. In Section~III we present elastic scattering
results for proton scattering from a variety on nuclei based on
the Nijmegen and CD-Bonn potentials. In Section~IV we present a detailed
study on which off-shell regions of the NN t-matrix are sampled in
a calculation of the elastic scattering observables. 
This study is based on the
factorized $t\rho$ approximation to the full-folding optical potential
and is carried out at 200 MeV projectile energy. 
We end with concluding remarks in Section~V.

\section{Theoretical Framework for the Optical Potential}

\hspace*{10mm}
The transition amplitude for elastic scattering of a projectile from
a target nucleus is given as \cite{med2}
\begin{equation}
T_{el} = P U P + P U G_0(E) T_{el}, \label{eq:2.1}
\end{equation}
where $P$ is the projector on the ground state $|\Phi_A\rangle$ of
the target, $P = \frac{|\Phi_A\rangle \langle \Phi_A|}
{\langle \Phi_A| \Phi_A\rangle}$, $G_0(E)=(E-H_0 +i\varepsilon)^{-1}$,
and $U$ represents the optical potential.
For the scattering of a single particle projectile from an A-particle
target nucleus the free Hamiltonian is given by $H_0=h_0+H_A$, where 
$h_0$ is the kinetic energy operator for the projectile
and $H_A$
stands for the target Hamiltonian.  In the spirit of the spectator
expansion the target Hamiltonian is viewed as $H_A=h_i + \sum_{j\neq i}
v_{ij} +H^i$, where $h_i$ is the kinetic energy operator for the
$i$th target nucleon,  $v_{ij}$  the interaction between target
nucleon $i$ and the other target nucleons $j$, and $H^i$ is an
(A-1)-body operator containing all higher order effects.
In a mean field approximation $\sum_{j\neq i} v_{ij} \approx W_i$,
where $W_i$ is assumed to depend only on the $i$th particle
coordinate. In this present work we want to concentrate only on the
impulse approximation, which is a good approximation in the intermediate
energy regime (around 200~MeV projectile energy and higher), where the
influence of $W_i$ can be neglected \cite{med2}. Thus the propagator
$G_0(E)$ in the impulse approximation is given as
\begin{equation}
G_0(E) \approx g_i(E)=[(E-E^i)-h_0 -h_i  + i \varepsilon]^{-1}.
 \label{eq:2.2}
\end{equation}
Here $H^i$, having no explicit dependence on the $i$th particle,
is replaced by an average
energy $E^i$.  In the present calculations we set $E_i=0$.
In the energy regime considered in this work, the effect of a value of 
$E_i$ of the order of the separation energy of a nucleon from a nucleus
is negligible \cite{edff,swth}.

\hspace*{10mm}
The driving term of Eq.~(\ref{eq:2.1}) denotes the optical potential,
which in first order is given as
\begin{equation}
\langle {\bf k'}|\langle \Phi_A| PUP|\Phi_A\rangle |{\bf k}\rangle
\equiv {\hat U}({\bf k'},{\bf k}) = \sum_{i=n,p}\langle {\bf k'}|
\langle \Phi_A| {\hat \tau}_{0i}({\cal E})|\Phi_A \rangle |{\bf k}\rangle.
  \label{eq:2.3}
\end{equation} 
Here ${\bf k'}$ and ${\bf k}$ are the external momenta of the system,
${\hat \tau}_{0i}({\cal E})$ represents the NN transition operator
\begin{equation}
{\hat \tau}_{0i}(E) = v_{0i} + v_{0i} g_i(E) {\hat \tau}_{0i}(E) ,
\label{eq:2.4}
\end{equation}
with $g_i(E)$ given in Eq.~(\ref{eq:2.2}) and $v_{0i}$ representing the
NN interaction. The sum over $i$ in
Eq.~({\ref{eq:2.3}) indicates the two different cases, namely when the
target nucleon is one of Z protons, and when it is one of N neutrons.
The energy ${\cal E}$ is the relative energy of the interacting 
two-nucleon system.
Inserting a complete set of momenta for the struck target nucleon before
and
after the collision and evaluating the momentum conserving
$\delta$-functions
gives as final expression for the full-folding optical potential
\cite{ff1,swth}
\begin{eqnarray}
\hat{U}({\bf q}, {\bf K})= \sum_{i=n,p} \int d^3P &&
\;\eta({\bf P},{\bf q}, {\bf K})\;\hat{\tau}_{0i}({\bf q},
\frac{1}{2}(\frac{A+1}{A}{\bf K}-{\bf P}), {\cal E}) \nonumber \\
&&\rho_i({\bf P}-\frac{A-1}{A}\frac{{\bf q}}{2},
{\bf P}+\frac{A-1}{A}\frac{{\bf q}}{2}) \label{eq:2.5}.
\end{eqnarray}
Here the arguments of the NN amplitude $\hat{\tau}_{0i}$ are
${\bf q}={\bf k}'-{\bf k}={\bf k_{NN}}'-{\bf k_{NN}}$ and
$\frac{1}{2}({\bf k_{NN}}'+{\bf k_{NN}}) =
\frac{1}{2}(\frac{A+1}{A}{\bf K}-{\bf P})$, where
\begin{equation}
{\bf k'_{NN}}=\frac{1}{2}({\bf k}' - ({\bf P}-{\frac{\bf q}{2}} -
\frac{\bf K}{A}) \label{eq:2.5a}
\end{equation}
 and
\begin{equation}
{\bf k_{NN}}=\frac{1}{2}({\bf k} - ({\bf P}+{\frac{\bf q}{2}} -
\frac{\bf K}{A}) \label{eq:2.5b}
\end{equation}
 are the nonrelativistic final and initial
nuclear momentum in the zero momentum frame of the NN system,
and ${\bf K}=\frac{1}{2}({\bf k'}+{\bf k})$.
The factor $\eta({\bf P},{\bf q}, {\bf K})$ is the M\o ller
factor for the frame transformation~\cite{Joachain}, and
$\rho_i$ represents the density matrix of the target for either protons
or neutrons.
Evaluating the propagator $g_i(E)$ of Eq.~(\ref{eq:2.2}) in
the nucleon-nucleus (NA) center of mass frame yields for the
energy argument ${\cal E}$ of the NN amplitude
$\hat{\tau}_{0i}$ of Eq.~(\ref{eq:2.5})
\begin{equation}
{\cal E} = E_{NA} - \frac{(\frac{A-1}{A}{\bf K}+{\bf P})^2}{4m_N}.
\end{equation}
Here $E_{NA}$ is the total energy in the NA center of mass frame
and $m_N$ is the nucleon mass.

\hspace*{10mm}
The expression for the  optical potential as given in
Eq.~(\ref{eq:2.5}) shows that the evaluation of the full-folding
integral requires the NN t-matrix fully off-shell as well as at positive
energies from $E_{NA}$ to negative energies \cite{hugo1,hugo2,edff}.

\section{Proton Elastic Scattering Observables}

\hspace*{10mm}
In this paper the study of elastic scattering of protons from spin
zero target nuclei at energies between 100 and 200~MeV incident
projectile energy is strictly first order based on the impulse
approximation. The full-folding optical potentials are calculated
according to Eq.~(\ref{eq:2.5}). The details of the calculations are
given in Refs.~\cite{ff1,edff}. As a model for the density matrix 
for the target nucleus we employ a Dirac-Hartree (DH) calculation~\cite{DH}.
The Fourier transform of the vector density,
$\rho({\bf r}',{\bf r})$, serves as our
non-relativistic single particle density~\cite{ff1}. The crucial
ingredient under investigation here is the fully off-shell NN t-matrix.
The calculations presented here employ NN t-matrices based on 
two different potentials given by the Nijmegen group 
\cite{nijm} and the charge-dependent Bonn potential \cite{cdbonn}.
All three potentials are fitted to describe the Nijmegen data base with
a $\chi^2$ per datum $\sim$~1. An essential difference between the two
Nijmegen models is the presence of a momentum dependent, nonlocal term
in the central piece of the NijmI potential, whereas the NijmII model is
strictly local. Both Nijmegen potentials have a $\chi^2$ per datum =~1.03 
with respect to both, the neutron-proton and the proton-proton
data base. The CD-Bonn potential is nonlocal due to the structure of the
relativistic meson-nucleon vertices. An additional nonlocality is
contained due to the socalled minimal relativity factors $\sqrt{m/E}$,
which are necessary to maintain the relativistic unitarity condition.
The CD-Bonn potential also describes the Nijmegen data base with
a $\chi^2$ per datum =~1.03. All three potential models describe the
Nijmegen data base with the same high degree of accuracy, thus the NN
t-matrices can be considered on-shell equivalent. From their different
theoretical derivation it can be expected that they have different
extrapolations off-shell.

\hspace*{10mm}
When calculating $\hat{U}({\bf q}, {\bf K})$ as given in
Eq.~(\ref{eq:2.5}), it is to be understood that all spin summations
are carried out. This reduces the required NN t-matrix elements to a
spin independent component (corresponding to the Wolfenstein amplitude A) 
and a spin-orbit component (corresponding to the Wolfenstein amplitude
C). Since we are assuming that we have spin saturated nuclei, the
components of the NN t-matrix depending on the spin of the struck
target nucleon vanish. The Coulomb interaction between the projectile
and the target is included using the exact formulation of
Ref.~\cite{coul}.

\hspace*{10mm}
At first we want to concentrate on proton scattering from different
target nuclei at 200~MeV projectile energy. In Fig.~1 we display the 
differential cross section $d\sigma/d\Omega$, the analyzing power $A_y$,
and the spin rotation function $Q$ for elastic proton scattering from
$^{16}$O. The solid line represents a calculation based employing the
CD-Bonn t-matrix as input, the dashed line is based on the one derived
from the NijmI potential and the dash-dotted line the one derived from
the NijmII model. All three calculations are remarkably close to each
other, and all three fail to describe the dips in the analyzing power.
The same statement is true for proton scattering from $^{40}$Ca at 
200~MeV, which is displayed in Fig.~2. In Fig.~3 we show the elastic
scattering observables for proton scattering from $^{208}$Pb at 
200~MeV. Again, all three NN potential models give nearly identical
results, however the spin observables are described slightly better
for $^{208}$Pb.

\hspace*{10mm}
At lower energies the scattering observables may exhibit a somewhat
larger sensitivity to the energy dependence of the NN t-matrix due to
the closer proximity of the deuteron pole and the virtual $^1$S$_0$
state. In order to study the sensitivity of the NA scattering observables 
to different NN t-matrices at lower energies we show in Fig.~4 
the observables for proton scattering from $^{40}$Ca  at 160~MeV and
in Fig.~5 the ones for proton scattering from $^{16}$O at 135~MeV.
Again, all three potential models lead to nearly identical results.

\hspace*{10mm}
We do not want to carry out further studies at lower energies, since it
is well known that the impulse approximation alone is not adequate to
describe the scattering observables at lower energies
\cite{med2,hugo1,edff}. We prefer to pursue further
investigations to find out why expected off-shell differences in the
potential models are not visible in the elastic NA observables.

\section{Investigation of Off-Shell Differences}

\hspace*{10mm}
In the full-folding optical potential as given in Eq.~(\ref{eq:2.5})
the energy of propagation in the NN t-matrix is coupled to the
integration variable. This makes it difficult to access effects
resulting from the off-shell structure of the NN t-matrices separately.
For this reason, we prefer to carry out the following study using the
optimum factorized form of the optical potential, which has been shown
to be quite a good approximation to the full-folding expression at
projectile energies of 200~MeV and higher \cite{edff}. The optimum
factorized form is characterized by two approximations. First, the
energy $\cal E$ of the NN t-matrix in Eq.~(\ref{eq:2.5}) is fixed at
half the projectile energy (in the laboratory frame)
\begin{equation}
{\cal E} \equiv E_0 = \frac{1}{2} \frac{k^2_{lab}}{2m_N}=
\frac{1}{2}\frac{{(\frac{A+1}{A}k_0)}^2}{2m_N}.\label{eq:3.1}
\end{equation}
Here $k_{lab}$ and $k_0$ are the on-shell momenta in the
laboratory and NA system respectively, and
$m_N$ is the mass of a nucleon. Second, the NN t-matrix and
the  M\o ller factor are expanded
in  ${\bf P}$ around a fixed value ${\bf P_0}$, determined by the 
requirement that the contribution of the first derivative term is
minimized. For elastic scattering the contribution vanishes if 
${\bf P_0}$ is chosen to be zero \cite{pttw,ernst}. With these
assumptions, the expression for the optical potential in the optimum
factorized form is given as
\begin{equation}
{\hat U}_{fac}({\bf q},{\bf K})=\sum_{i=p,n}
\;\eta({\bf q},{\bf K})\;\hat{\tau}_{0i}
\left({\bf q},{{A+1}\over{2A}}{\bf K} ,E_0 \right)\; 
\rho_{i}\left(q \right). \label{eq:4.2}
\end{equation}
In this form the non-local character of the optical potential is
solely determined by the off-shell NN t-matrix and the
 M\o ller factor. If we now consider the
integral equation for elastic NA scattering as given in Eq.~(\ref{eq:2.1}),
we see that only the second term in the right hand side of
Eq.~(\ref{eq:2.1}) contains the integration over the optical potential. 
The driving term, ${\hat U}_{fac}({\bf k'_0},{\bf k_0},E)$ contains the
NN t-matrix evaluated at the fixed momenta ${\bf k'_0}$ and
 ${\bf k_0}$, multiplied
with the density profile $\rho_i(q)$. In this case the momentum
vectors  ${\bf q}$ and ${\bf K}$ are 
${\bf q}={\bf k'}_0 - {\bf k}_0$ and ${\bf K}=\frac{1}{2}
({\bf k'}_0+{\bf k}_0)$.

\hspace*{10mm}
In order to study off-shell effects, we define the following 
quantity
\begin{equation}
B({\bf k'}_0,{\bf k}_0,E)= \lim_{\epsilon \rightarrow 0}
\int_0^{\infty} {d^3{\bf k''}}\;\frac{ {\hat U}_{fac}
({\bf k'}_0,{\bf k''},E) \;
T({\bf k''},{\bf k}_0,E)}{E -E(k'')+ i\epsilon}, \label{eq:4.3}
\end{equation}
where $T({\bf k''},{\bf k}_0,E)$ is the solution of
Eq.~({\ref{eq:2.1}}), obtained using the optical potential in the
factorized form.  Here $B({\bf k'}_0,{\bf k}_0,E)$ represents the
integral on the right-hand side of Eq.~({\ref{eq:2.1}}), and thus
the quantity in which the optical potential $U$ enters off-shell when
calculating $T_{el}$.
Since the nuclear density in momentum space is a
function strongly peaked for small momenta, we may conjecture that the
density will dominate the fall-off behavior of 
${\hat U}_{fac} ({\bf k'}_0,{\bf k''},E)$ for large values of the integration
variable ${\bf k''}$. To investigate this more closely, we write 
Eq.~(\ref{eq:4.3}) as
\begin{equation}
B({\bf k'}_0,{\bf k}_0,E)= \lim_{\epsilon \rightarrow 0}
\int d\Omega'' \int_0^{k_{max}}
dk'' \; k''^2 \frac{ {\hat U}_{fac}({\bf k'}_0,{\bf k''},E) \;
T({\bf k''},{\bf k}_0,E)}{E -E(k'')+i\epsilon}, \label{eq:4.4}
\end{equation} 
and study the behavior of $B({\bf k'}_0,{\bf k}_0,E)$ as a function of
$k_{max}$. Since $B({\bf k'}_0,{\bf k}_0,E)$ depends on  vector variables,
we actually have $B(k_0,k_0,\theta,E)$, where $\theta$ is the
angle between ${\bf k'_0}$ and ${\bf k_0}$. In Fig.~6 we show
the real part $Re \; B(k_0,k_0,\theta,E)$ 
for different values of $k_{max}$ for neutron
scattering from $^{16}$O at 200~MeV projectile energy, and in 
Fig.~7 for neutron scattering from $^{90}$Zr at the same energy.
We see that in the case of $^{16}$O an integration up to $k_{max} =
k_0+1.0$~fm$^{-1}$ is already sufficient to obtain the full result.
In the case of $^{90}$Zr one only needs to integrate to 
$k_{max} = k_0+0.5$~fm$^{-1}$ to have a result identical to the
complete integral. In both cases $k_0 \sim 3$~fm$^{-1}$. When considering
the imaginary part of $B(k_0,k_0,\theta,E)$ we arrive at the same
conclusion.  We carried out
similar tests at different energies and arrived essentially at the same
values for $k_{max}$ for the two different nuclei. Assuming that the 
nuclear density is responsible for the fast fall-off of the optical
potential as function of ${\bf k'}$, this finding is not surprising.
From Figs.~6 and 7 we also see that for a heavier nucleus the
contribution beyond the on-shell value $k_0$ is much less than for a
light nucleus. Again, this is not too surprising, when one recalls the
functional form of the nuclear density profiles. The density profile
$\rho_p (q)$ for the proton distribution of $^{16}$O has its first
minimum at $q \sim 2$~fm$^{-1}$, whereas the proton distribution of 
$^{90}$Zr has its first minimum at $q \sim 1$~fm$^{-1}$.

\hspace*{10mm}
In order to verify that the functional form of the density
is the limiting factor for the range of the integration, we
identify in Eq.~(\ref{eq:4.4}) $T({\bf k''},{\bf k}_0,E)$ as well 
as ${\hat U}_{fac}({\bf k_0},
{\bf k''},E)$ with the density $\rho(|{\bf k''}-{\bf k_0}|)=\sum_{i=p.n}
\rho_i(|{\bf k''}-{\bf k_0}|)$ to obtain
\begin{equation}
B'({\bf k'_0},{\bf k_0},E)= \lim_{\epsilon \rightarrow 0}
\int d\Omega'' \int_0^{k_{max}}
dk'' \; k''^2 \frac{ \rho(|{\bf k''}-{\bf k_0}|)
\rho(|{\bf k''}-{\bf k'_0}|)}{E -E(k'')+i\epsilon}, 
\label{eq:4.5}
\end{equation} 
and repeat the above study, namely consider $B'(k_0,k_0,\theta,E)$ 
as a function of $k_{max}$. In Fig.~8 we plot the real part
$Re \;B'(k_0,k_0,\theta,E)$ using different
integration ranges. The result is similar to the one in Fig.~6 and 7.
Considering  Fig.~8 the upper bound for the
integral can be constrained to  $k_0 +1.5~$fm$^{-1}$. Since 
the integral in Eq.~(\ref{eq:4.5}) is symmetric about the on-shell
value, the lower bound of integration can be constrained to
 $k_0 -1.5~$fm$^{-1}$.

\hspace*{10mm}
After having found that only a limited region of the NN t-matrix enters
a calculation of NA elastic scattering observables, we need to project
this region, namely $k_0 \pm 1.5~$fm$^{-1}$, on the NN t-matrices and
see if the t-matrices employed in our calculations differ in this
restricted region. Thus we show first in Fig.~9 the real part
of the off-shell Wolfenstein amplitude A,
 $Re \;A(k'_{NN},k_{NN},E_0)$, at
200 MeV obtained from the NijmI potential as function of $k_{NN}$ 
and $k'_{NN}$. Here  the angle between ${\bf k_{NN}}$ and 
${\bf k'_{NN}}$ is chosen to be zero. The value of the on-shell momentum is 
located at $k_{NN}=k'_{NN}=1.55$~fm$^{-1}$.  It should be noted that the
values of $Re \;A$ in the plotted domain range between 
0.6 and -3~MeVfm$^3$.
Since the study of the integration bounds in the integral
$B(k_0,k_0,\theta,E)$ was carried out using momenta defined in the NA
system, we use Eqs.~(\ref{eq:2.5a}) and (\ref{eq:2.5b}) to transform
the bounds to momenta given in the NN system. As a reminder, 
since we work in the optimum factorized form, the momentum ${\bf P}$ in
Eqs.~(\ref{eq:2.5a}) and (\ref{eq:2.5b}) is zero. Using these
transformations, which are explicitly given as
${\bf k'_{NN}}= \frac{1}{4} [(1/A+3){\bf k'_0}+(1/A-1){\bf k''}]$
and ${\bf k_{NN}}= \frac{1}{4} [(1/A-1){\bf k'_0}+(1/A+3){\bf
k''}]$, we obtain the `skew box' given in Fig.~9 as region of
the NN t-matrix whose values  enter the NA scattering equation.

\hspace*{10mm}
Next, we display in Fig.~10 the difference between the real parts
of the Wolfenstein amplitudes $Re \;A(k'_{NN},k_{NN},E_0)$
 given
by the NijmI and CD-Bonn potentials, again as function of $k_{NN}$,
$k'_{NN}$ and the angle between the two vectors being zero.
First we notice that within the plotted region the off-shell differences
between the two amplitudes is relatively small. Only for $k_{NN}=
k'_{NN} \sim 5$~fm$^{-1}$ there is a difference larger than
0.2~MeVfm$^3$. Again, the on-shell value is located at
$k_{NN}=k'_{NN}=1.55$~fm$^{-1}$. The region which enters a calculation of
NA scattering observables is again indicated by a `skew box'. Within this
box there are essentially no differences between the amplitudes.
The largest difference is located in the upper right corner of
the `skew box' almost opposite the on-shell point, and is about 6\% 
of the total value of $Re \;A$. 

\hspace*{10mm}
In Fig.~11 we show the difference between the real parts of the
Wolfenstein amplitudes derived from the NijmI
and NijmII potentials. These two  amplitudes show off-shell differences of
1~MeVfm$^3$ and larger for values of $k_{NN}=k'_{NN} \sim 4$~fm$^{-1}$.
However, in the region which is sampled by NA scattering calculations
(`skew box') both amplitudes are nearly identical in the lower left
half around the on-shell value. Larger differences between those two
potentials are located in the uppler right corner furthest away from
the on-shell value.  For the other Wolfenstein
amplitude,  which enters our calculations of elastic NA scattering 
observables, we obtain similar conclusions.
A close inspection of the scattering observables for the light nuclei
$^{16}$O and $^{40}$Ca in Figs.~1 and 2 shows that the dash-dotted curves
representing the calculations with the NijmII model can be distinguished
from the other two curves, especially at larger angles. However,
the differences in the observables are still quite small, indicating
the calculations are dominated by the area around the on-shell value.

Thus we can see, that although the Wolfenstein amplitudes A and C
 derived from the different NN
potentials under study exhibit differences for large off-shell momenta,
the off-shell region which is sampled in NA elastic scattering
calculations is restricted to an area close to the on-shell value
and thus does not  probe those far off-shell regions where the larger
differences occur.

\section{Summary and Conclusion}

\hspace*{10mm}
In this paper we addressed the question if nucleon-nucleus 
elastic scattering observables are sensitive to different off-shell
structures of NN transition amplitudes derived from realistic NN potentials.
Our study is based on the recently developed potential models NijmI and
NijmII by the Nijmegen group  \cite{nijm} and the charge dependent Bonn
potential \cite{cdbonn}. All three potentials models describe the
Nijmegen NN data base with a $\chi^2$ per datum =1.03. Thus the 
transition matrices derived from these models can be considered on-shell
equivalent. The Wolfenstein amplitudes, which enter our NA calculations,
show considerable differences for large off-shell momenta. However,
these differences are not visible in the NA elastic scattering
observables.

\hspace*{10mm}
We calculated elastic scattering observables for proton scattering from
$^{16}$O, $^{40}$Ca, and $^{208}$Pb in the energy regime between 
100 and 200~MeV projectile energy. Here we calculated the full-folding
integral for the first order optical potential using the impulse
approximation within the framework of the spectator expansion of
multiple scattering theory. In addition to the NN t-matrices from the
three above mentioned potential models our optical potentials employ a
Dirac-Hartree model for the nuclear density matrix. Recoil and frame
transformation factors are implemented in the calculation in their
complete form. We find that the elastic scattering observables based on
the three different potential models are almost identical. A very
similar result has been obtained in Ref.~\cite{hugo1}. This work
employs different density matrices and is based on the Paris potential
and inversion potentials which are constructed to be phase-shift
equivalent to the Paris potential as well as to the experimentally
extracted phase shifts.

\hspace*{10mm}
In order to better understand our numerical results, we study the regions
of the NN t-matrices, which are sampled in a calculation of NA elastic
scattering observables within the off-shell $t\rho$ or optimum
factorized approximation to the full-folding optical potential. In this
approximation the off-shell character of the optical potential is solely
determined by the off-shell NN t-matrix. This feature allows us to
determine, which region of off-shell momenta for a fixed energy slice of
the NN t-matrix enter the calculation. Our investigation of the
rescattering term of the Lippmann-Schwinger equation shows, that the
off-shell dependence of the optical potential is limited by the nuclear
density, which in momentum space is a strongly peaked function for small
momenta. It is well known that the heavier the nucleus becomes, 
the stronger is that forward peaking. This
property of the nuclear density prevents far off-shell momenta of the NN
t-matrix from entering the optical potential and thus the NA scattering
observables. 
The coincidence of the calculations based on the
different realistic NN potentials strongly indicates that only off-shell
momenta close to the on-shell value of the NN t-matrix are relevant
for NA scattering. In this region the different potentials still give
very similar results for the NN t-matrix. 

\hspace*{10mm}
Comparing our calculations of elastic scattering observables to
experimental data, we still find some systematic inabilities of the
first order full-folding optical potential to describe certain details
for the NA scattering data in the considered energy regime. However, 
we find the
limitations of the first order optical potential cannot be attributed
to uncertainties associated with the off-shell behavior of the realistic
NN t-matrices employed.

\vfill
\acknowledgments
The authors want to thank W. Gl\"ockle for many stimulating, helpful and
critical discussion during this project.

This work was performed in part under the auspices of the U.~S.
Department of Energy including contract No. DE-FG02-93ER40756 with
Ohio University.  One of us (D.H.) would like to thank the Deutsche
Forschungsgemeinschaft for their support.
We thank the Ohio Supercomputer Center (OSC) for
the use of their facilities under Grant
No.~PHS206
as well as the National Energy Research Supercomputer Center
(NERSC) for the use of their facilities
under the FY1997 Massively Parallel Processing Access Program.



\newpage

\noindent

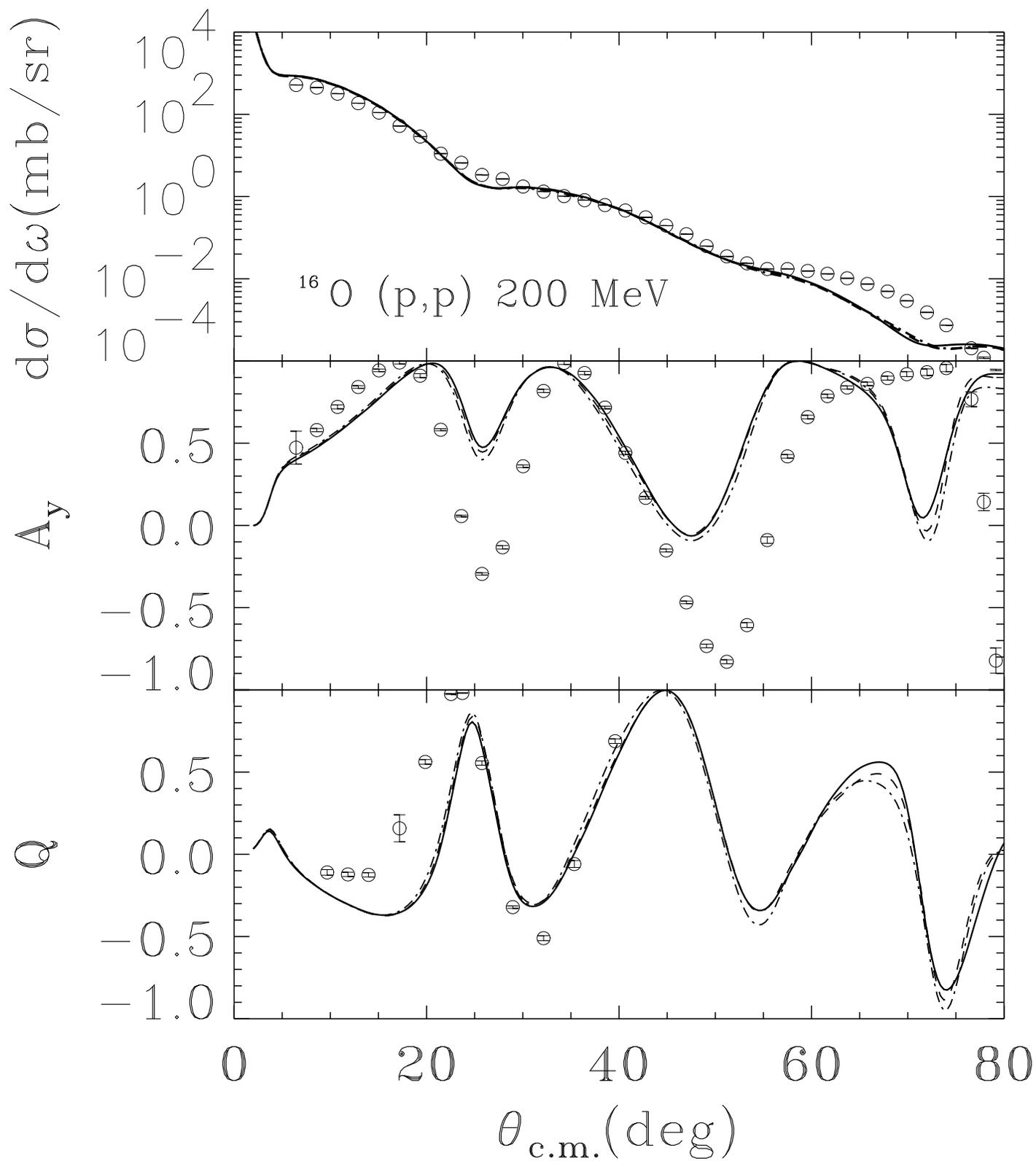
\begin{figure}
\caption{ The angular distribution of the differential cross-section
         ($\frac{d\sigma}{d\Omega }$), analyzing power ($A_y$) and
         spin rotation function ($Q$) are shown for elastic proton
         scattering from $^{16}$O at 200 MeV laboratory energy.
The solid line represents the calculation performed with a
first-order full-folding optical potential based on the DH density
\protect\cite{DH} and the CD-Bonn model \protect\cite{cdbonn}. The 
dashed line uses the NijmI model instead, the dash-dotted line the
NijmII model \protect\cite{nijm}. The data are taken from
Ref.~\protect\cite{O200}. \label{fig1}}
\end{figure}

\noindent
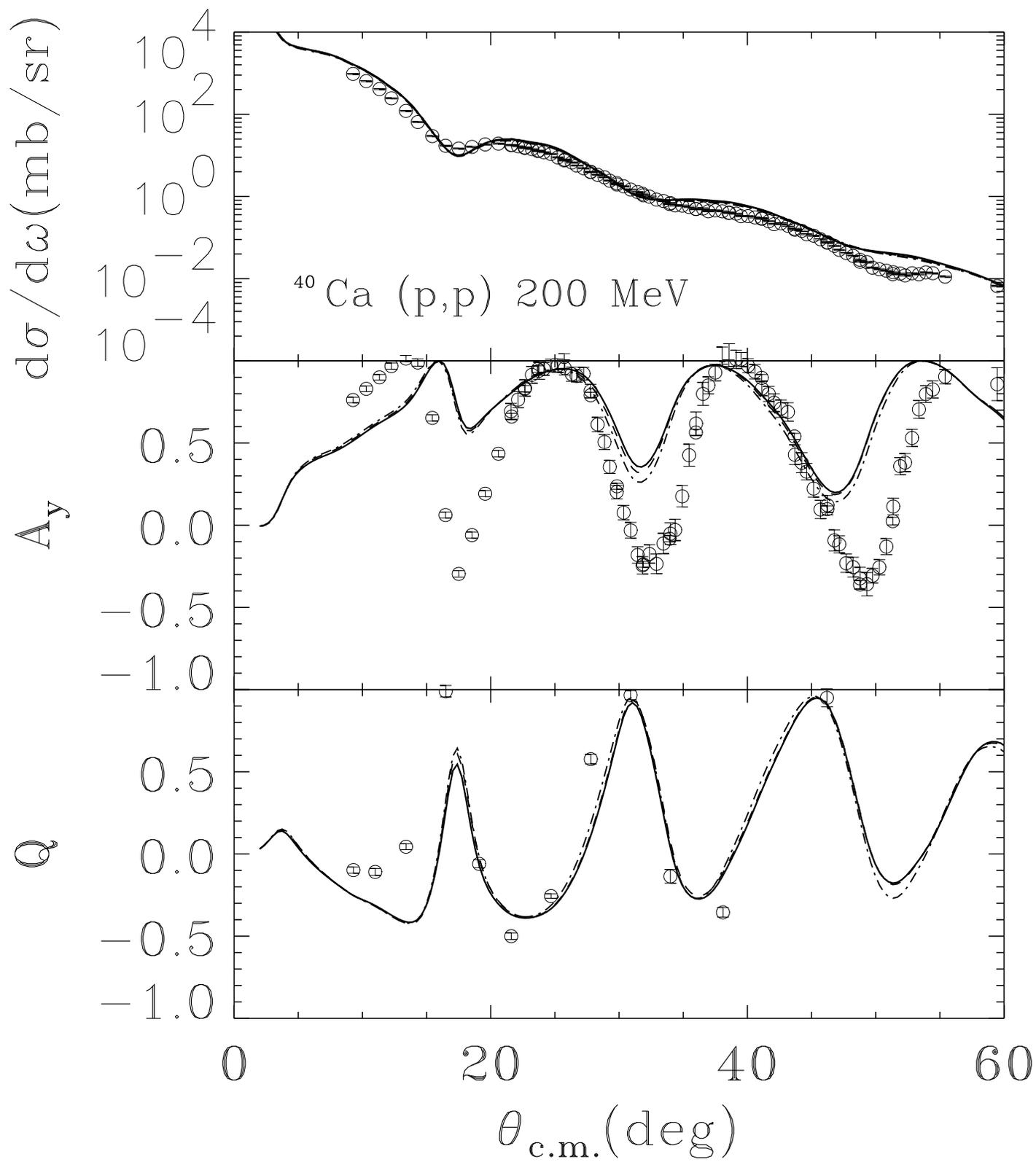
\begin{figure}
\caption{Same as Fig.~1, except that the target nucleus is $^{40}$Ca.
The data are taken from Ref.~\protect\cite{ca200}. \label{fig2}}
\end{figure}

\noindent
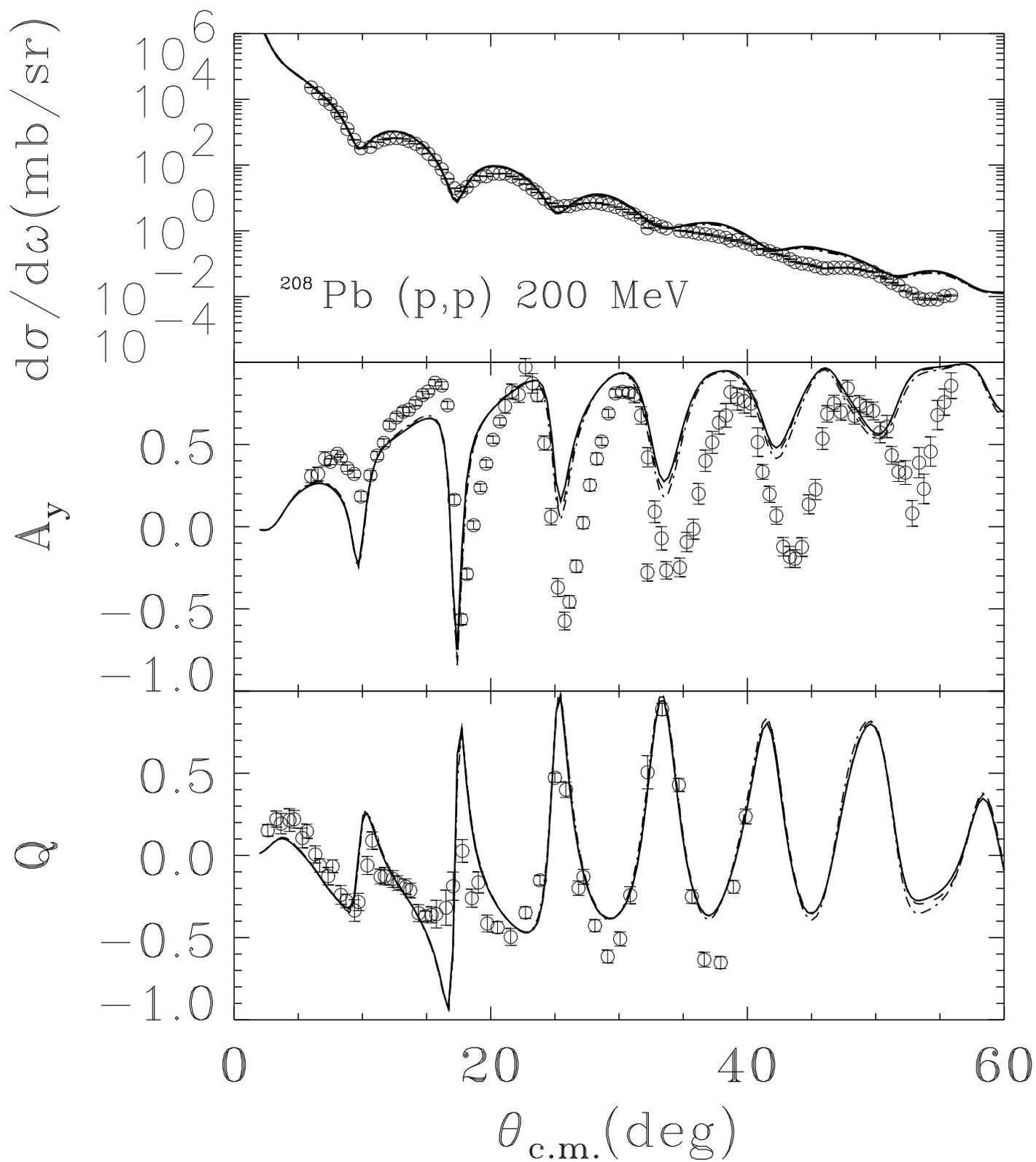
\begin{figure}
\caption{Same as Fig.~1, except that the target nucleus is $^{208}$Pb.
The data are taken from Ref.~\protect\cite{ca200}. \label{fig3}}
\end{figure}

\noindent
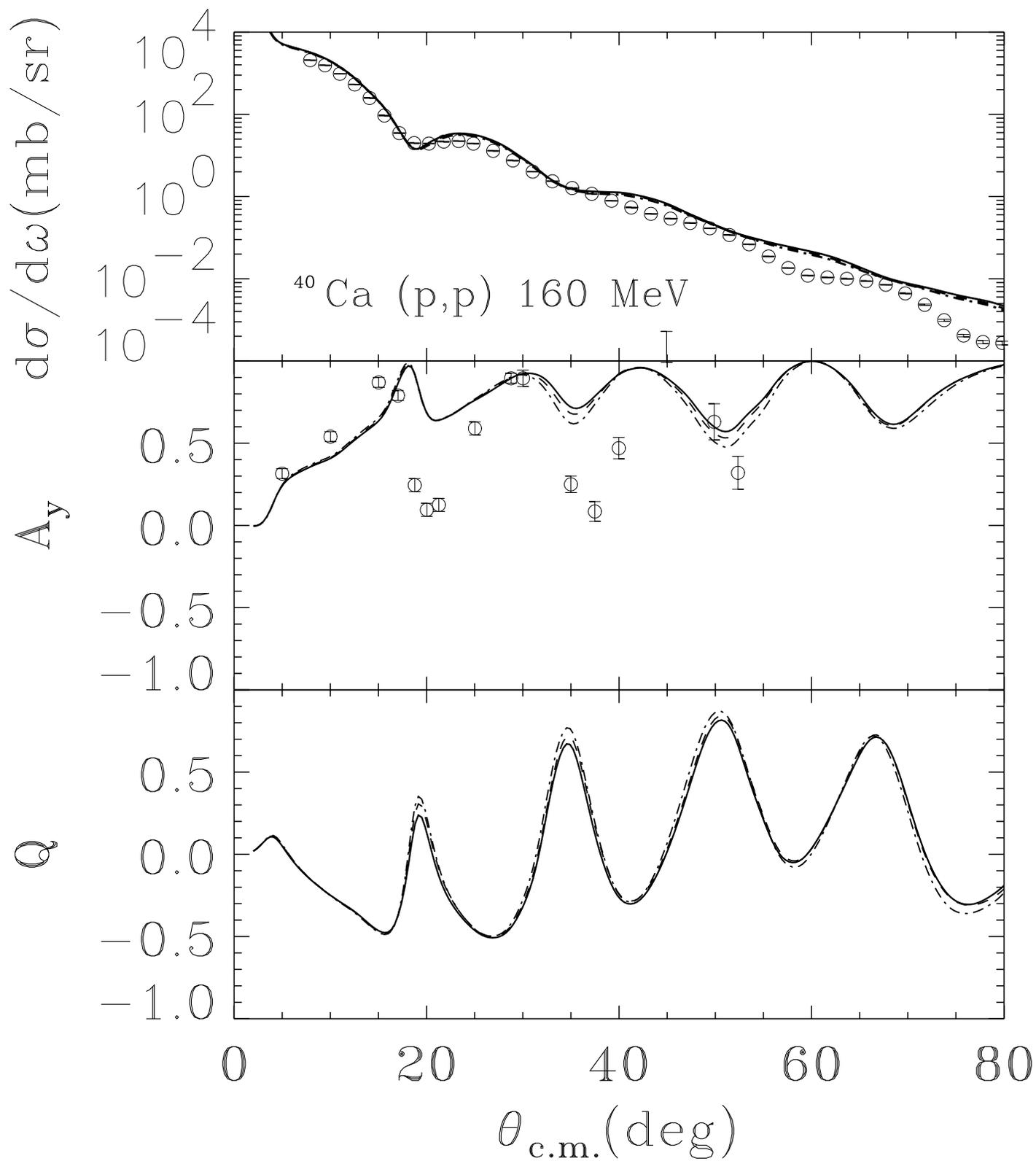
\begin{figure}
\caption{Same as Fig.~2, except that the projectile energy is
160~MeV. The data are taken from Ref.~\protect\cite{ca160}.
\label{fig4}}
\end{figure}

\noindent
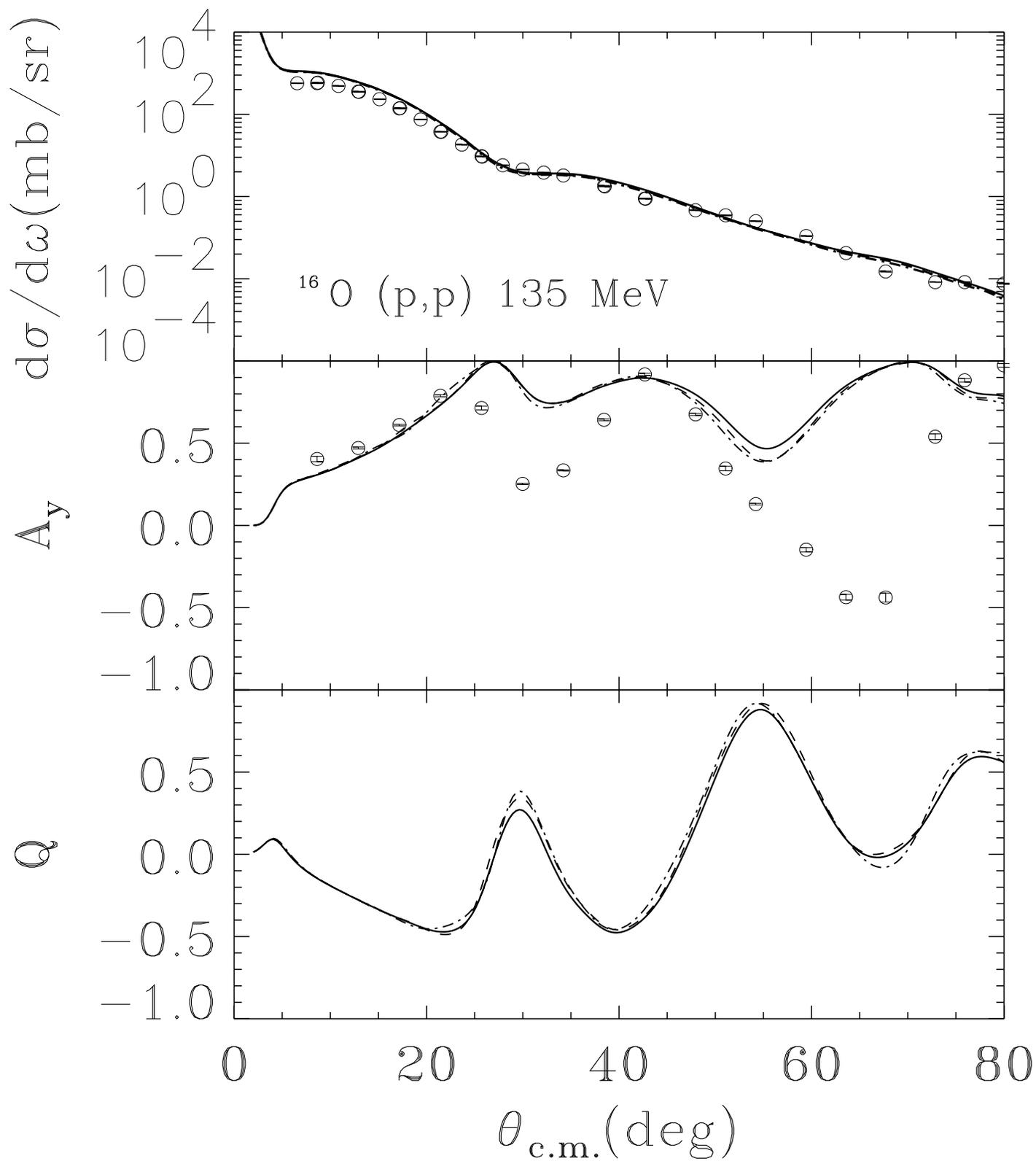
\begin{figure}
\caption{Same as Fig.~1, except that the projectile energy is
135~MeV. The data are taken from  Ref.~\protect\cite{o135}.
\label{fig5}}
\end{figure}

\noindent
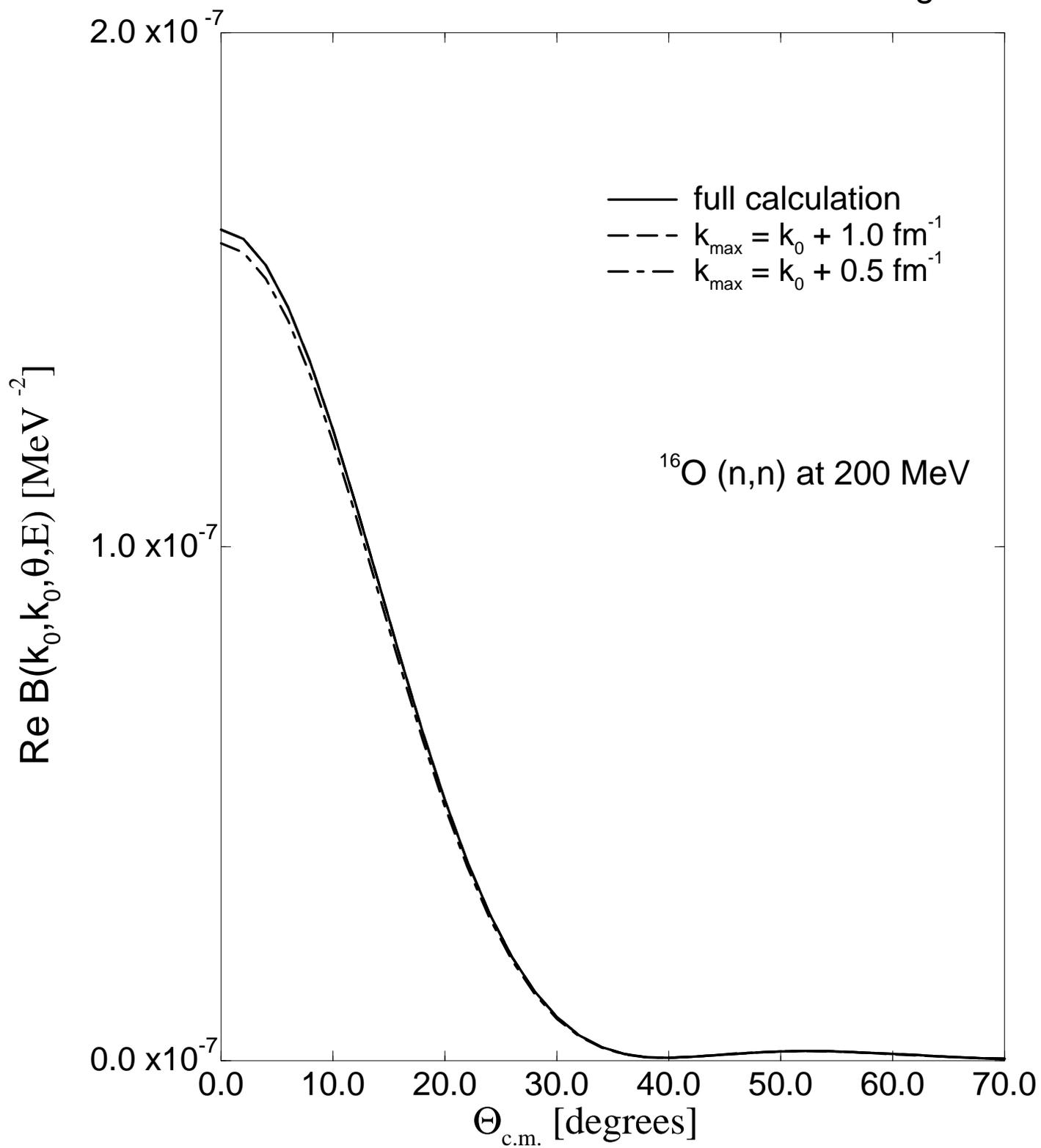
\begin{figure}
\caption{ The real part of the function $B(k_0,k_0,\theta,E)$ as
defined in Eq.~(\protect\ref{eq:4.4}) for neutron scattering from
$^{16}$O at 200 MeV laboratory energy.
The solid line represents the full calculation, 
the dashed line, which coincides with the solid line, represents
the calculation with an upper limit $k_{max}$ in the
k-integration of $k_{max}=k_0 +1.0$~fm$^{-1}$, while the
dot-dashed line represents a calculation using  $k_{max}=k_0
+0.5$~fm$^{-1}$.
\label{fig6}}
\end{figure}

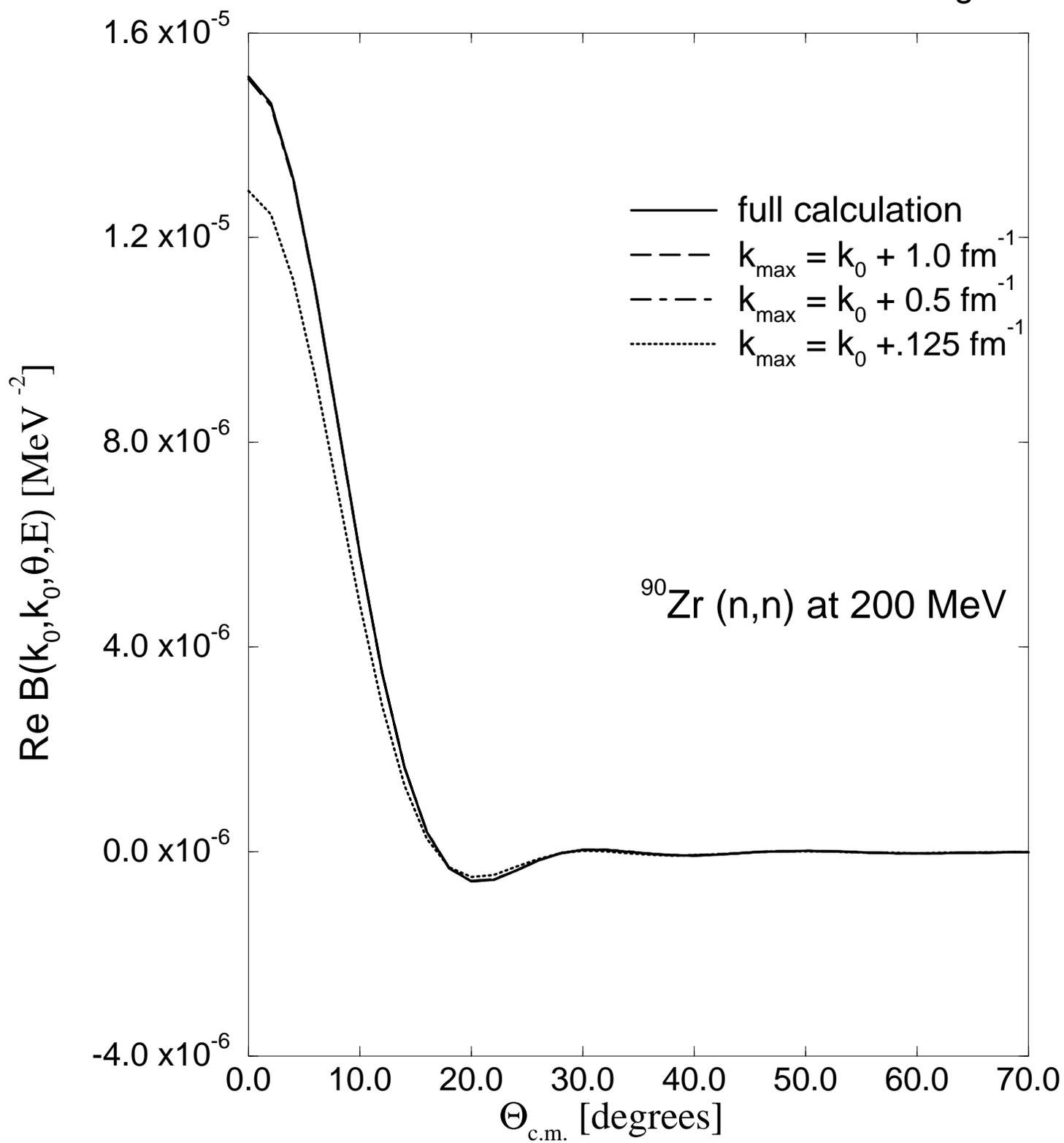
\begin{figure}
\caption{Same as Fig.{\protect\ref{fig6}} that the nucleus is
$^{90}$Zr. The solid line represents the full calculation, 
the dashed line, which coincides with the solid line, represents
the calculation with an upper limit $k_{max}$ in the
k-integration of $k_{max}=k_0 +1.0$~fm$^{-1}$, while the
dot-dashed line represents a calculation using  $k_{max}=k_0
+0.5$~fm$^{-1}$. For the dotted line $k_{max}=k_0 +0.125$~fm$^{-1}$
was used.
\label{fig7}}
\end{figure}

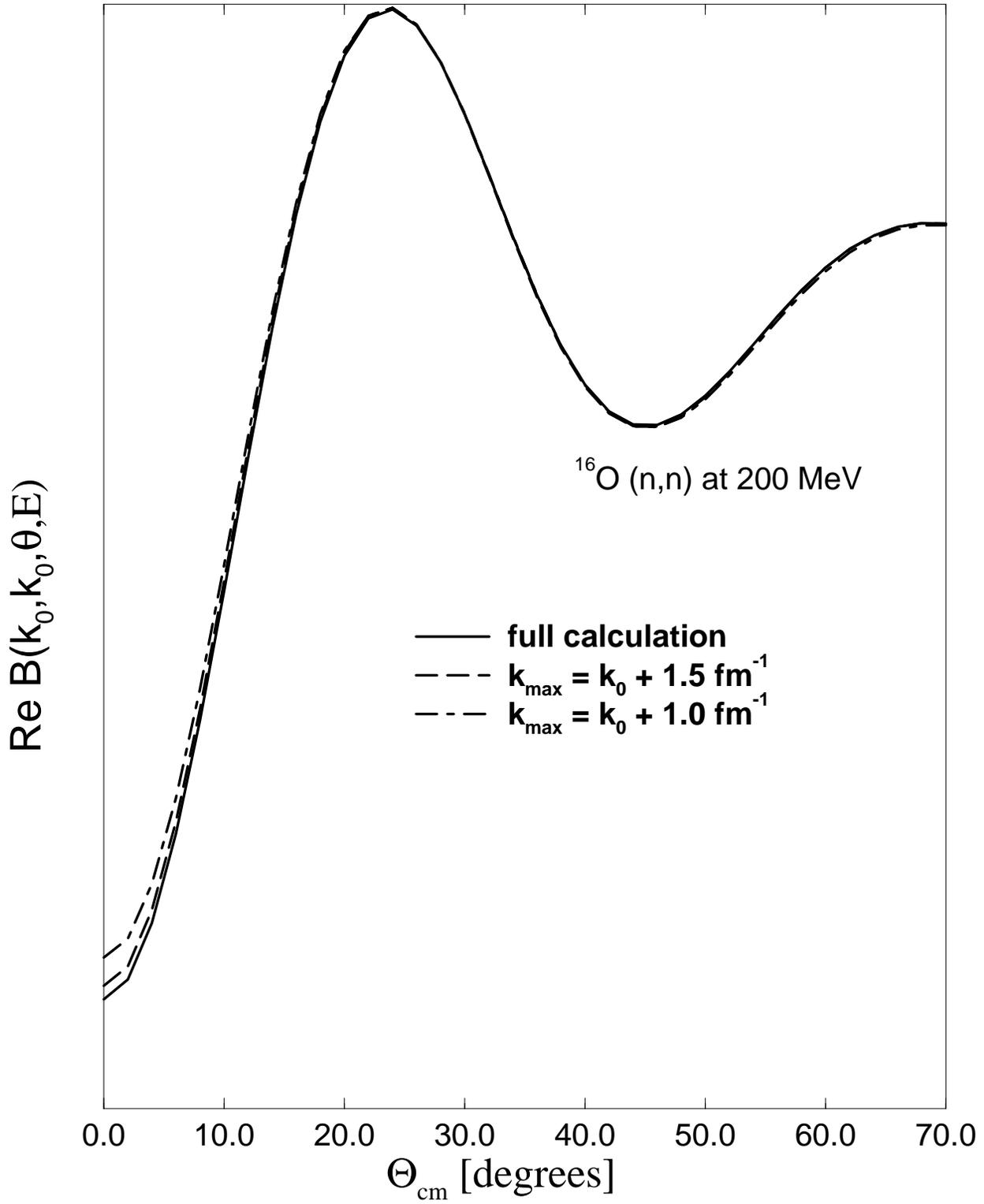
\begin{figure}
\caption{ The real part of the function $B'(k_0,k_0,\theta,E)$ as
defined in Eq.~(\protect\ref{eq:4.5}) for $^{16}$O at 200 MeV laboratory energy.
The solid line represents the full calculation. The the dash-dotted line
represents the calculation with an upper limit $k_{max}$ in the 
k-integration of $k_{max}=k_0 +1.5$~fm$^{-1}$, while the 
dashed line represents a calculation using  $k_{max}=k_0 
+1.0$~fm$^{-1}$. The values for Re $B'$ are given in arbitrary units.
\label{fig8}}
\end{figure}

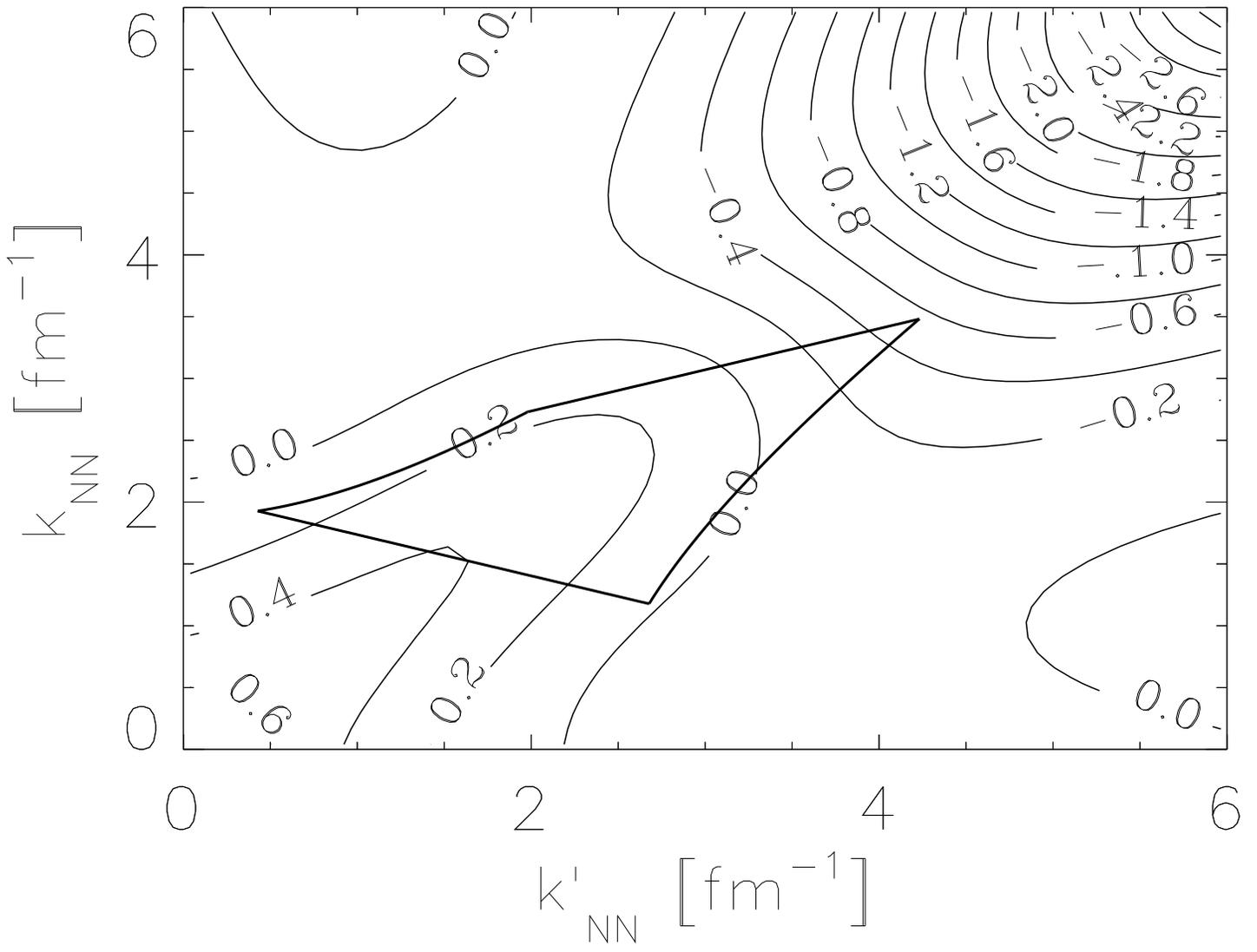
\begin{figure}
\caption{ The Wolfenstein amplitude $Re \;A(k'_{NN},k_{NN},E_0)$ 
for np scattering at 200 MeV obtained from the NijmI potential
{\protect\cite{nijm}}. The angle between ${\bf k_{NN}}$ and
${\bf k'_{NN}}$ is chosen to be zero. The contour lines represent
steps of 0.2 MeVfm$^3$. The `skew box' represents the region of
momenta, which is accessed by a calculation of the NA scattering
observables as described in the text.
\label{fig9}}
\end{figure}

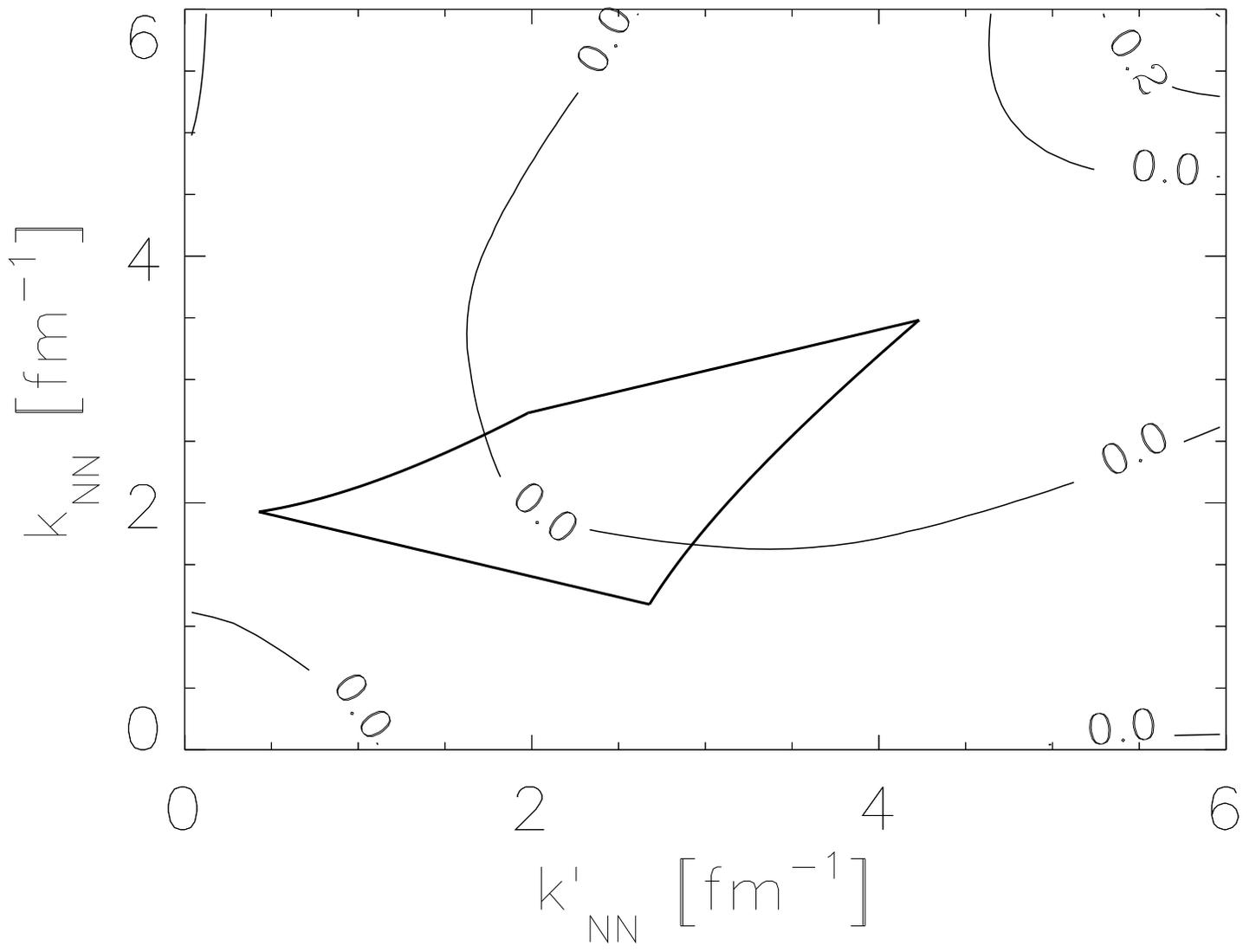
\begin{figure}
\caption{ The difference of the Wolfenstein amplitudes
$Re \;A(k'_{NN},k_{NN},E_0)$ obtained by subtraction the 
amplitude obtained from the CD-Bonn potential {\protect\cite{cdbonn}}
from the one obtained from the NijmI potential
{\protect\cite{nijm}}. The angle between ${\bf k_{NN}}$ and
${\bf k'_{NN}}$ is chosen to be zero. The contour lines represent
steps of 0.2 MeVfm$^3$.
\label{fig10}}
\end{figure}
 
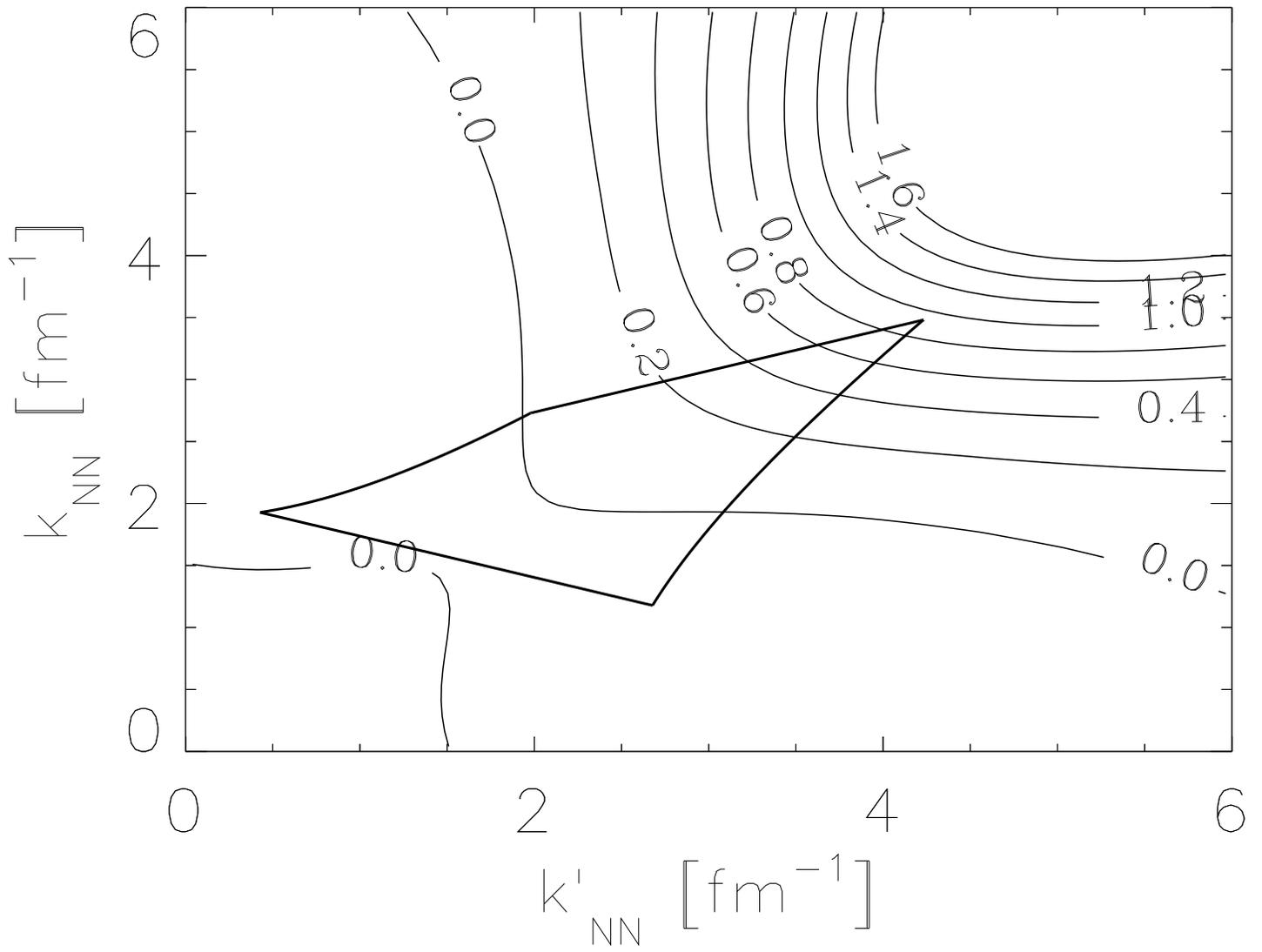
\begin{figure}
\caption{Same as Fig.~10, except that here the real part of the
Wolfenstein amplitude A obtained from the NijmII potential is subtracted
from the one obtained from the NijmI potential.
\label{fig11}}
\end{figure}

\end{document}